\documentclass[preprint,12pt]{elsarticle}

\usepackage{graphics}
\usepackage{epsfig}

\usepackage{amssymb}

\usepackage[dvips]{color}
\usepackage{colordvi}
\usepackage{bm}
\usepackage{graphics}
\def\beq{\begin{equation}}
\def\be{\begin{eqnarray}}
\def\eeq{\end{equation}}
\def\ee{\end{eqnarray}}

\def\lsim{\buildrel < \over {_{\sim}}}
\def\gsim{\buildrel > \over {_{\sim}}}

\journal{Physics Letters B}

\begin{document}

\begin{frontmatter}



\title{Shear viscosity of $\beta$-stable nuclear matter}


\author[label1,label2]{Omar Benhar}
\author[label2]{Arianna Carbone}

\address[label1]{INFN, Sezione di Roma, I-00185 Roma, Italy}
\address[label2]{Dipartimento di Fisica, ``Sapienza'' Universit\`a di Roma, 
I-00185 Roma, Italy}

\begin{abstract}
Viscosity plays a critical role in determining the stability of rotating neutron stars. 
We report the results of a calculation of the shear viscosity of $\beta$~-~stable matter, carried
out using an effective interaction based on a state-of-the-art nucleon-nucleon potential and the 
formalism of correlated basis functions.
Within our approach the equation of state, determining the proton fraction, and the nucleon-nucleon 
scattering probability are consistently obtained from the same dynamical model.
The results show that, while the neutron contribution to the viscosity is always dominant, above nuclear 
saturation density the electron contribution becomes appreciable.
\end{abstract}


\end{frontmatter}



\section{Introduction}
\label{intro}

The quantitative description of transport properties of nuclear matter is relevant to the understanding
of a variety of neutron star properties. Thermal conductivity is one of the driving factors of the cooling
process, while electrical conductivity is relevant to the ohmic dissipation of magnetic
fields in the star interior. In rotating stars, a crucial role is also played by viscosity, that determines the
possible onset of the gravitational-wave driven instabilities first predicted by Chandrasekhar in the
1970s \cite{Chand1,Chand2}.

Gravitational radiation (GR) is emitted when a non-radial oscillation mode
 of the star is excited by an internal or external perturbation. In non rotating neutron stars the
emission of GR is a dissipative process, leading to the damping of the oscillation, while in rotating
 stars the effect can be quite different. At the end of the 1970s, Friedman and Schutz \cite{FS} proved
that, due to the mechanism discussed in Refs.\cite{Chand1,Chand2}, all perfect fluid rotating stars are in fact
unstable.
More recently, Andersson \cite{And} and Friedman and Morsink \cite{FM} demonstrated that all the so called 
{\em r} modes,
i.e. the oscillations of rotating stars whose restoring force is the Coriolis force, are driven unstable by GR in 
all perfect fluid stars.
On the other hand, if matter in the star interior does not behave as a perfect fluid, dissipative processes, such as 
viscosity,
can damp the modes responsible of the instability, or even suppress them completely. Hence, knowledge
of the viscosity of neutron star matter is required to determine whether a mode is stable or unstable

The main limitation  of  most analyses of the damping of neutron-star oscillations \cite{cut_lind}
lies in the lack of consistency between the dynamical models used to obtain the equation of state (EOS),
describing the equilibrium properties of the star, and those employed to describe transport properties.
In their seminal paper, Cutler and Lindblom \cite{cut_lind} combined a variety of EOS, resulting from different
theoretical approaches, with the pioneering estimates of the shear viscosity coefficient of neutron star
matter obtained in the 1970s by Flowers and Itoh, who used  the the 
Landau-Abrikosov-Khalatnikov \cite{baym-pethick,ak}
formalism and the neutron-neutron collision probability estimated from the 
{\em measured} scattering phase shifts \cite{FI1,FI2}.

Nuclear many body theory provides a consistent framework to obtain the {\em in medium}
nucleon-nucleon (NN) cross section and the transport coefficients of nuclear matter from realistic NN potentials,
using either the $G$-matrix \cite{wambach} or the CBF \cite{BV}
formalism. In both approaches one can define a well behaved effective
interaction, suitable for use in standard perturbation theory in the Fermi gas basis
and allowing for a {\em unified} treatment of equilibrium and non equilibrium properties
\cite{wambach,BV,shannon,gof4,BF09}.



The CBF effective interaction has been employed to carry out a  calculation
of the shear viscosity of pure neutron matter \cite{BV}. However, a more realistic model of neutron star matter 
must allow for the presence of protons and electrons. As matter density increases, the electron 
chemical potential may also exceed the muon rest mass, making the appearance of muons 
energetically favorable.

In this  Letter, we discuss the generalization of the approach of Ref.\cite{BV} to the case of $\beta$-stable matter
consisting of neutrons, protons and electrons.

\section{Formalism}
\label{formalism}

The application of the Abrikosov-Khalatnikov \cite{ak} formalism  to the calculation of the shear viscosity 
of matter consisting of neutrons, protons and electrons in $\beta$-equilibrium was first developed by 
Flowers and Itoh \cite{FI1,FI2}. Within their approach, the NN scattering rate
due to strong interactions is modeled using the measured free space cross section, thus neglecting 
all modifications caused by the presence of the nuclear medium.

In Refs.\cite{FI1,FI2}, the calculation of the transport coefficients of $\beta$-stable
matter is carried through a straightforward generalization of the case of pure neutron matter. 
In a multicomponent system the Boltzmann-Landau equation takes the form
\begin{equation}
\frac{\partial n_\alpha}{\partial t}+\frac{\partial n_\alpha}{\partial{\bf r}}\cdot\frac{\partial\epsilon_{{\bf p}\alpha}}
{\partial{\bf p}}-\frac{\partial n_\alpha}{\partial{\bf p}}\cdot\frac{\partial\epsilon_{{\bf p}\alpha}}{\partial{\bf r}}= \sum_{\beta} I_{\alpha \beta} \ ,
\label{BLE}
\end{equation}
where $n_\alpha = n_{{\bf p}\alpha}({\bf r},t)$ denotes the distribution of quasiparticles 
of type $\alpha$ ($\alpha = n, p, e)$, carrying momentum {\bf p} and energy $\epsilon_{{\bf p}\alpha}$ . The form of the 
collision term in the right hand side of the above equation clearly  
shows that, in principle, all binary collisions, involving both like and unlike quasiparticles, must be taken into account. 

The shear viscosity, defined as the coefficient of the momentum flux tensor appearing in the left hand 
side of Eq.(\ref{BLE}), can be written as \cite{FI2}
\beq
\eta=\eta_n+\eta_p+\eta_e\ ,
\label{eta:sum}
\eeq
the contribution associated with quasiparticles of type $\alpha$  being given by \cite{FI1,FI2,BS1,BS2} 
\begin{equation}
\eta_\alpha=\frac{1}{5}\rho_\alpha m_\alpha^\star v_{F\alpha}^2\tau_\alpha \frac{2}{\pi^2(1-\ell_{\alpha\alpha})}
C(\ell_{\alpha\alpha}) \ .
\label{etatot}
\end{equation}
In the above equation, $\rho_\alpha$, $m_\alpha^\star$ and $v_{F\alpha}$ denote the density, effective mass and 
Fermi velocity, respectively, while the quasiparticle lifetime $\tau_\alpha$ is given by
\begin{equation}
\tau_\alpha=\frac{4\pi^4}{m_\alpha^\star T^2 \ \sum_\beta m_\beta^{*2} \langle W_{\alpha\beta} \rangle} \ ,
\label{def:tau}
\end{equation}
where $T$ is the temperature, and
\begin{equation}
\ell_{\alpha\alpha}=\frac{\sum_\beta \langle W_{\alpha\beta} L_{\alpha\beta}^\alpha \rangle }{\sum_\beta
 \langle W_{\alpha\beta} L_{\alpha\beta} \rangle} \ .
\label{lambda}
\end{equation}
In Eq.(\ref{lambda}), $W_{\alpha \beta}$ denotes the probability of collisions between quasiparticles 
of type $\alpha$ and $\beta$. In the low temperature limit, underlying the Landau-Abrikosov-Khalatnikov 
approach, scattering processes can only involve quasiparticles
carrying momenta close to the Fermi momentum. As a consequence, at fixed baryon density, 
$\rho= \rho_p + \rho_p$, and proton fraction $x=\rho_p/\rho$,  $W_{\alpha \beta}$ only depends on two angular variables, 
$\theta$ and $\phi$, and the averages in Eqs.(\ref{def:tau}) and (\ref{lambda}) are defined as
\beq
\langle F \rangle = \int  \frac{d\Omega}{4 \pi} \ F(\theta,\phi) \ .
\eeq 
The quantities $L_{\alpha\beta}^\alpha$ and $L_{\alpha\beta}$ are also functions of the angles $\theta$ and $\phi$.
Their explicit expressions, as well as that of the factor $C(\ell_{\alpha\alpha})$, are given in Refs.~\cite{FI1,FI2}. 

\section{Results}
\label{res}

For any given baryon density, the calculation of the shear viscosity requires the knowledge of the proton fraction
$x$, determined by the conditions of $\beta$-equilibrium and charge neutrality
\beq
\mu_n - \mu_p = \mu_e \ \ \ , \ \ \ \rho_p = \rho_e \ , 
\label{conditions}
\eeq
where $\mu_\alpha$ denotes the chemical potential of quasiparticles of type $\alpha$. 

In this work, the proton and neutron chemical potentials have been computed within the Hartree-Fock approximation
\beq
\mu_\alpha = \epsilon_\alpha(p_{F \alpha}) \ ,
\eeq
where the Fermi momentum is given by $p_{F \alpha} = (3 \pi^2 \rho_\alpha)^{1/3}$, 
using the single particle spectrum obtained from the CBF effective interaction of Ref.\cite{BV} in the Hartree-Fock
approximation
\beq
\epsilon_\alpha(p) = x_\alpha \left\{ \frac{p^2}{2m} +  
\rho \sum_\beta x_\beta \int d^3x  
\left[ \langle v_{\rm eff} \rangle_D - \langle v_{\rm eff} \rangle_E \  \ell(p_{F \alpha} x) \ {\rm e}^{i {\bf p}\cdot{\bf x}}
\right] \right\} \ .
\label{ep1}
\eeq
In the above equation, $m$ denotes the nucleon mass, $\ell(x) = 3(\sin x - x \cos x)/x^3$ and the spin averaged
direct and exchange matrix elements of the effective interaction are given by
\beq
\langle v_{\rm eff} \rangle_D = \frac{1}{2} \ \sum_{{\sigma_\alpha}{\sigma_\beta}} \ 
\langle \alpha \beta | v_{\rm eff} | \alpha \beta \rangle \ \ , \ \ 
\langle v_{\rm eff} \rangle_E = \frac{1}{2} \ \sum_{{\sigma_\alpha}{\sigma_\beta}} \ 
\langle  \alpha \beta | v_{\rm eff} |  \beta \alpha \rangle \ .
\label{ep2}
\eeq
The effective interaction $v_{\rm eff}$  
is based on a truncated version of the NN potential referred to as Argonne $v_{18}$ \cite{av18},  providing an 
excellent fit of deuteron properties and the full Nijmegen phase shift database. 
It also includes the effects of interactions involving three- and many-nucleon forces, described according to the approach 
originally proposed in Ref.\cite{LagPan}. The EOS of symmetric nuclear matter and pure
neutron matter obtained using the dynamical model of Ref.\cite{BV} turn out to be in fairly good agreement with 
the results of the state-of-the-art calculations of Ref.\cite{APR}, carried out using the full Argonne $v_{18}$ potential.
\begin{figure}
\begin{center}
\includegraphics[scale=0.49]{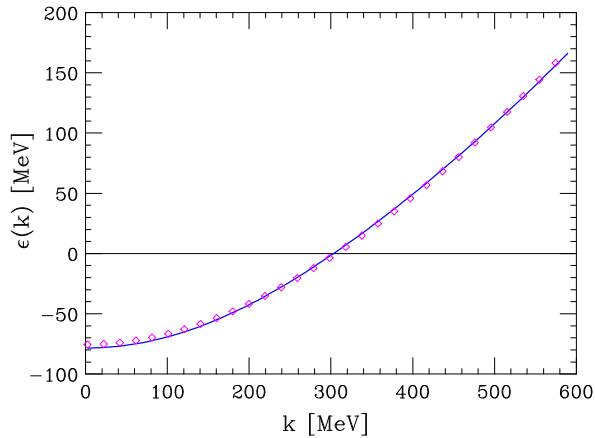}
\caption{Solid line: single particle spectrum evaluated using Eq.(\ref{ep1}) and the effective interaction
of Ref.\cite{BV}. The diamonds show the results of Ref.\cite{bob}, obtained using the FHNC
approach and a realistic nuclear hamiltonian.}
\label{sp}
\end{center}
\end{figure}

In Fig. \ref{sp} the single particle spectrum of symmetric nuclear matter at equilibrium density obtained from 
Eq.(\ref{ep1}) is compared to the results of Ref.\cite{bob}, carried out 
within the Fermi Hyper-Netted Chain (FHNC) approach using a realistic nuclear hamiltonian.

\begin{figure}
\begin{center}
\includegraphics[scale=0.5]{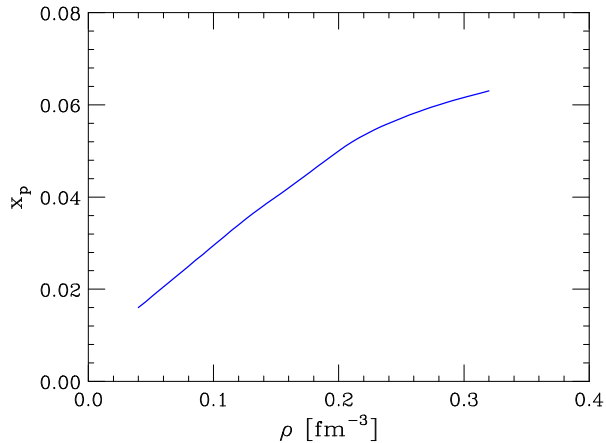}
\caption{Proton fraction of charge neutral $\beta$-stable matter consisting of neutrons, protons and electrons, 
obtained from the effective interaction of Ref.\cite{BV}, as a function of baryon density.}
\label{pfrac}
\end{center}
\end{figure}

Figure \ref{pfrac} shows the baryon density dependence of the proton fraction resulting from the numerical solution 
of Eqs.(\ref{conditions}), carried out assuming that electrons can be described using the single particle spectrum 
of the relativistic Fermi gas. 
The effective masses $m^*_\alpha$ appearing in Eqs.(\ref{etatot}) and (\ref{def:tau}) can be 
readily obtained from the single particle energies through
\begin{equation}
\frac{1}{m_\alpha^\star}=\frac{1}{p}\frac{d\epsilon(p)}{dp} \ .
\end{equation}

The scattering probabilities employed in our calculations take into account both strong and electromagnetic
interaction. Nuclear interactions contributing to $W_{nn}$=$W_{pp}$ and $W_{np}$=$W_{pn}$ have
been described within the dynamical model of  Ref.\cite{BV}, while the calculation of $W_{ep}$=$W_{pe}$ , 
$W_{en}$=$W_{ne}$ and the 
electromagnetic part  of $W_{pp}$ have been carried following Refs.\cite{FI1,FI2}.
\begin{figure}
\begin{center}
\includegraphics[scale=0.5]{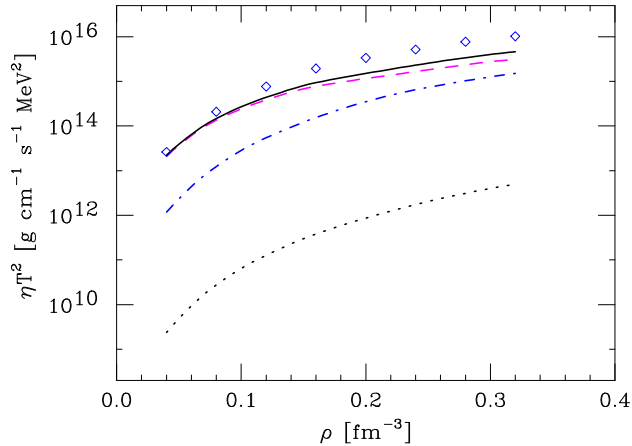}
\caption{Baryon density dependence of the quantity $\eta_\alpha T^2$ for 
protons (dotted line), electrons (dot-dashed line) and neutrons (dashed line) in charge 
neutral $\beta$-stable matter. The solid line corresponds to the total $\eta T^2$ (see Eq.(\ref{eta:sum})). 
For comparison, the diamonds show $\eta T^2$ of pure neutron matter, corresponding to 
$x_p=0$.}
\label{etat2}
\end{center}
\end{figure}

Figure \ref{etat2} shows the temperature-independent quantities $\eta_\alpha T^2$, as well 
as $\eta T^2$, plotted as a function of baryon density.  It appears that the proton viscosity $\eta_p$ is always very 
small, due to their low density and mobility. On the other hand, in spite of the fact that $\rho_p$=$\rho_e$, the 
electron viscosity $\eta_e$ turns out to be much higher, as electrons are ultra-relativistic. As expected, the 
dominant contribution is  $\eta_n$, as the neutron fraction is larger than 90\% over the whole range of baryon 
density. For $\rho \lsim \rho_0$, $\rho_0 =$ 0.16 fm $^{-3}$ being the equilibrium density of symmetric nuclear matter, 
the total viscosity can be identified with the neutron contribution. The electron contribution
becomes barely visible only at larger density.

The main difference between the case of a multi-component fluid and that of pure neutron matter, discussed in Ref.\cite{BV},
 lies in the larger number of channels available in collision processes, which implies a  shorter quasiparticle lifetime,
leading in turn to a lower viscosity. This feature is illustrated by the diamonds of Fig. \ref{etat2}, showing the energy 
dependence of the viscosity of pure neutron matter, corresponding to $x_p=0$.  

To gauge the dependence of our results on the composition of matter, we have repeated the calculations 
using a larger proton fraction, obtained by multiplying the results shown in Fig. \ref{pfrac} by a factor two. 
Note that changing the proton fraction amounts to using a different model of nuclear dynamics.  Larger proton fractions 
correspond to a larger contribution of the symmetry term to nuclear matter energy. 
\begin{figure}
\begin{center}
\includegraphics[scale=0.49]{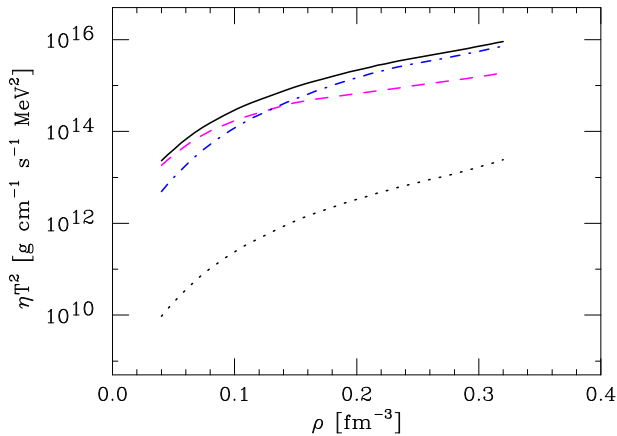}
\caption{Same as in Fig. \ref{etat2}, but with the proton (and electron) fractions 
 of Fig. \ref{pfrac} increased by a factor two.}
\label{eta3}
\end{center}
\end{figure}

As shown in Fig. \ref{eta3}, in this case the electron viscosity exceeds neutron viscosity
 at values  of baryon density $\rho \sim 0.2$ fm$^{-3}$, just above nuclear matter saturation density,
  corresponding to an electron  fraction of $\sim$ 8\%.



\section{Conclusions}
\label{summary}

The results of our work show that inclusion of medium effects on 
NN scattering leads to a sizable suppression of the collision probability, 
producing in turn an enhancement of the shear viscosity of nuclear matter.

The role played by screening of the bare NN interaction, mainly due to short range correlations, 
was already pointed out in Ref. \cite{BV} for the case of pure neutron matter. Our study of
charge-neutral $\beta$-stable matter consisting of neutrons, protons and electrons, 
carried out using the same effective interaction, suggests that, while the proton contribution 
to viscosity can be safely neglected, the electron contribution is appreciable. In the case of 
dynamical models predicting a larger proton fraction, typically $x_p \gsim 8 \%$ at $\rho \sim \rho_0$,  
 it may in fact become dominant at densities exceeding nuclear matter saturation
density. Overall, due to the availability of a larger number of reaction channels, the viscosity
of $\beta$-stable matter turns out to be lower than that of pure neutron matter. 

All the above considerations are based on the tenet that neutrons, protons and 
electrons behave as normal Fermi liquid. In the region of density and temperature in which neutrons 
become superfluid, their contribution to the viscosity vanishes and the electron contribution
takes over. 

In their study of the effect of viscosity on neutron star oscillations, the authors of Ref. \cite{cut_lind} 
took into account the onset of superfluidity writing the viscosity in the form
\beq
\eta_s = [ 1 - \Theta (\rho,T) ] \eta + \Theta (\rho,T) \eta_e \ ,
\eeq
where
\beq
\Theta (\rho,T) = 
\left\{ 
\begin{array}{cc}
0 & T > T_c \\
1 & T < T_c
\end{array}
\right. \ ,
\eeq
$T_c = T_c(\rho)$ being the critical temperature, taken from Ref.\cite{ao}. It has to be pointed out, however, that 
a fully consistent analysis of the stability of rotating neutron stars requires that the EOS, determining the proton fraction,  the viscosity coefficient and the critical temperature be all obtained from the same dynamical model.
The effective interaction approach appears to be ideally suited to pursue this project. 

Numerical calculations of the $^1S_0$ superfluid gap carried out using the effective interaction of 
Ref.\cite{BV} yield a critical temperature $T_c \sim 2 \times 10^{10}$ K at density $\rho \sim 0.04$ fm$^{-3}$, typical of the 
the neutron star inner crust \cite{salvi}, in fairly good agreement with the rsults of Ref.\cite{ao}. The extension of this study 
to the case of pairing in $^3P_2$ states is currently being carried out.

As a final remark, it is worth mentioning that a more realistic model of neutron star matter should include muons, 
whose appearance
is likely to be energetically favored at densities above nuclear saturation density. 
However, compared to electrons, muons have larger mass, and therefore lower mobility. As a consequence, 
taking into account their 
contribution to the viscosity is not expected to significantly affect the conclusions of our work.

\end{document}